\documentclass[aps,pra,final,twocolumn,floatfix,showpacs,10pt]{revtex4-1}
\usepackage{graphicx,amssymb,soul,color,amsmath,bm}
\usepackage{epstopdf,hyperref}
\usepackage{braket}
\usepackage{bbm} 
\usepackage{times}

\hypersetup{colorlinks=true,citecolor=blue}

\newcommand{\Eq}[1]{Eq.~(\ref{#1})}
\sethlcolor{yellow}

\begin{document}

\title{
Lieb-Liniger model with exponentially-decaying interactions:\\a continuous matrix product state study
}

\author{Juli\'an Rinc\'on}
\affiliation{Perimeter Institute for Theoretical Physics, Waterloo, Ontario, N2L 2Y5, Canada}

\author{Martin Ganahl}
\affiliation{Perimeter Institute for Theoretical Physics, Waterloo, Ontario, N2L 2Y5, Canada}

\author{Guifre Vidal} 
\affiliation{Perimeter Institute for Theoretical Physics, Waterloo, Ontario, N2L 2Y5, Canada}

\date{\today}

\begin{abstract}
The Lieb-Liniger model describes one-dimensional bosons interacting through a repulsive contact potential. In this work, we introduce an extended version of this model by replacing the contact potential with a decaying exponential. Using the recently developed continuous matrix product states techniques, we explore the ground state phase diagram of this model by examining the superfluid and density correlation functions. {At weak coupling superfluidity governs the ground state, in a similar way as in the Lieb-Liniger model.} However, at strong coupling {quasi-crystal} and super-Tonks-Girardeau regimes are also found, which are not present in the original Lieb-Liniger case. Therefore the presence of the exponentially-decaying potential leads to a superfluid/super-Tonks-Girardeau/{quasi-crystal} crossover, when tuning the coupling strength from weak to strong interactions. This corresponds to a Luttinger liquid parameter in the range $K \in (0, \infty)$; in contrast with the Lieb-Liniger model, where $K \in [1, \infty)$, and the screened long-range potential, where $K \in (0, 1]$.
\end{abstract}


\maketitle


\section{Introduction\label{sec:intro}}
Outstanding developments in the field of cold atoms in optical lattices have opened the path to the experimental design and manipulation of many-body quantum states~\cite{Bloch08}. Specifically, there have been proposals to simulate quantum field theories using cold atoms~\cite{Cirac10,Bermudez10}. These proposals are of high relevance, as they can access non-perturbative regimes of such theories. Regarding the case of long-range interacting theories, there have been exciting experimental realizations with polar molecules~\cite{LongRange} that have enabled the exploration of strongly correlated phases not stable with local interaction potentials.

The theoretical description of such many-body physics is, however, challenging. Remarkable techniques have been proposed and applied to a variety of lattice models of strongly correlated systems~\cite{Schulz93,Tsukamoto00,Cazalilla03,Cazalilla04,Arkhipov05,Inoue06,Buechler07,Astrakharchik07,Citro07,Astrakharchik08,Citro08,Roscilde10,Dalmonte10}. Within this variety of methods, tensor networks are a set of ansatze for many-body wave functions proposed to tackle non-perturbative problems in lattice models, and have been successfully used to study strongly correlated phenomena~\cite{verstraete_matrix_2008,mcculloch_from,schollwock_density-matrix_2011}. The continuous counterpart of tensor networks has also been subject of recent research. More concretely, continuous matrix product states (CMPS) have been proposed as a variational ansatz to describe quantum field theories in $1+1$ dimensions~\cite{Verstraete10}. Another relevant proposal to study critical systems is the generalization to the continuum~\cite{JuthoMera} of the multiscale entanglement renormalization ansatz~\cite{Vidal07}.

Applications of CMPS to the study of many-body systems directly in the continuum include: ground state properties~\cite{Verstraete10} and excitations~\cite{Draxler13} of the Lieb-Liniger (LL) model, free massive Dirac fermions~\cite{Jutho10}, the $N$-flavor Gross-Neveu model~\cite{Jutho10}, and two-species of bosonic~\cite{Quijandria14} and fermionic~\cite{Chung15} systems. In particular for the bosonic case, such applications have been focused on continuous models with contact interactions. Studies of the long-range case have been explored before, using other techniques such as bosonization~\cite{Schulz93,Tsukamoto00,Cazalilla04,Inoue06,Citro07,Dalmonte10}, numerical computations~\cite{Arkhipov05,Buechler07,Astrakharchik07,Astrakharchik08,Citro08,Roscilde10}, and perturbation theory~\cite{Cazalilla03}. Although these approaches have certainly shed light on the physics of long-range interacting bosons, most methods have to be restricted to small values of coupling constants or must resort to a discretization procedure. Indeed, interesting phases such as superfluidity, Wigner crystal, charge-density wave, and Tonks-Girardeau gas have been found and their respective crossovers/transitions have been discussed. However, to the best of our knowledge, there are no studies of bosons with long-range interactions, directly in the continuum, using approaches with no such restrictions on length scales.

In this work, we use CMPS to study a system of bosons with exponentially-decaying interactions. This system can be considered as an extension of the exactly solvable LL Hamiltonian~\cite{Lieb63}. Inspired by lattice Hamiltonian models we will refer to this bosonic system as the extended Lieb-Liniger (ELL) model. We will see below that this extended Hamiltonian is a minimal model that captures both the physics of the LL Hamiltonian and that of bosons interacting through the screened Coulomb potential. When described as a Luttinger liquid, we will show that the Luttinger parameter of the extended Hamiltonian contains values taken by the corresponding parameter in the cases mentioned above. This will imply that our extended model displays strongly correlated behavior of both fermionic and bosonic character depending on the values of its parameters.

The outline of this work is the following. In Sec.~\ref{sec:model} we define the ELL model and show its connection to the LL Hamiltonian. Section~\ref{sec:ana} is devoted to a scaling analysis of the ELL model and its connection to long-range interacting bosons. The numerical results based on the variational CMPS will be presented and analyzed in Sec.~\ref{sec:num}. We summarize the main results in Sec.~\ref{sec:con}. Technical details of the CMPS are discussed in an Appendix.

\section{Hamiltonian Model\label{sec:model}}
In this work we will focus on a model of bosonic particles interacting via an exponentially-decaying density-density term in $1 + 1$ dimensions. Its Hamiltonian reads
\begin{equation}
\begin{split}
H &= \frac{1}{2m} \int dx\, \partial_x\psi^\dagger(x) \partial_x\psi(x) - \mu\int dx\, \psi^\dagger(x) \psi(x)\\
 &+ \frac{g}{2} \int dx\, dy \left(\frac{\eta}{2}\, e^{-\eta|x-y|}\right) \psi^\dagger(x) \psi^\dagger(y) \psi(y) \psi(x),
\label{eq:realH}
\end{split}
\end{equation}
where $\psi(x)$ and $\psi^\dagger(x)$ represent bosonic field operators that annihilate and create a particle at point $x$, respectively. The symbol $\partial_x$ stands for the partial derivate with respect to $x$, $\partial/\partial x$. The parameter $\mu > 0$ is the chemical potential, $m$ is the mass of the bosons that we set henceforth to $1/2$, $g \eta$ defines the interaction strength of the potential, and $\eta$ is a characteristic length that controls the range of the interaction.

Hamiltonian~(\ref{eq:realH}) is a particular case of a broader class of Hamiltonians that can be written down as
\begin{equation}
\begin{split}
H &= \frac{1}{2m} \int dx\, \partial_x\psi^\dagger(x) \partial_x\psi(x) - \mu\int dx\, \psi^\dagger(x) \psi(x)\\
 &+ \frac{1}{2} \int dx\, dy\, w(x-y) \psi^\dagger(x) \psi^\dagger(y) \psi(y) \psi(x),
\label{eq:H}
\end{split}
\end{equation}
where the interaction potential has been written as an arbitrary function $w(x-y)$ of the distance between $x$ and $y$. The long distance behavior of Hamiltonian~(\ref{eq:H}) was studied in Ref.~\onlinecite{DelMaestro10} invoking Luttinger liquid theory~\cite{Cazalilla11,GogolinBook,GiamarchiBook}. A Luttinger liquid is characterized by two independent parameters: the velocity of the excitations and the Luttinger parameter $K$. (Luttinger liquid theory is equivalent to a conformal field theory of free compactified massless scalar bosons~\cite{DiFrancescoBook}, with a compactification radius that depends on $K$.) This parameter controls the decay of the correlations and which particular phase governs the ground state.

We define the density-density and superfluid correlations as 
\begin{equation} 
\begin{split} 
C(x) &\doteq \frac{\langle \rho(x)\rho(0)\rangle}{\rho_0^2} - 1,\\
S(x) &\doteq \frac{\langle \psi^\dagger(x)\psi(0)\rangle}{\rho_0},
\end{split} 
\end{equation} 
where the density is defined as $\rho(x) \doteq \psi^\dagger(x)\psi(x)$ and its expectation value by the ground state is 
given by $\rho_0 = \langle \psi^\dagger(0)\psi(0) \rangle$. In Luttinger liquid theory the asymptotic expressions for these correlations at long distances, $\rho_0 x \gg 1$, are given by~\cite{Cazalilla04,Cazalilla11}
\begin{equation} 
\begin{split} 
C(\rho_0 x) &\approx -\frac{K}{2\pi ^ 2} \frac{1}{(\rho_0 x)^2} + A_1 \frac{\cos(2\pi\rho_0 x)}{(\rho_0 x)^{2K}},\\
S(\rho_0 x) &\approx \frac{1}{(\rho_0 x) ^{1/2K}}\left[ B_0 + B_1\frac{\cos(2\pi\rho_0 x)}{(\rho_0 x)^{2K}}\right].
\label{eq:corrs}
\end{split} 
\end{equation} 
In these expressions $A_1,\,B_0,$ and $B_1$ are coefficients that depend on the microscopic parameters in Hamiltonian~(\ref{eq:H}). Here we will follow the convention of calling the ground state of model~(\ref{eq:realH}) superfluid when $S(x)$ decays slower than $C(x)$. This will be satisfied if $K > 1/2$. Similarly, we will say that the ground state has charge order if $K < 1/2$, \emph{i.e.}, whenever $C(x)$ decays slower than $S(x)$. This convention stems from the fact that in one dimension there is no breaking of continuous symmetries; hence, algebraically-decaying correlations are the closest behavior to long-range order.

Two other instances of Hamiltonian~(\ref{eq:H}), previously analyzed in the literature, will be relevant to our discussion. The first one is the LL model~\cite{Lieb63}, which describes a system of bosons interacting through a contact potential weighted by the factor $g > 0$
\begin{equation}
w_{\rm LL}(x-y) = g\, \delta(x-y).
\label{eq:wLL}
\end{equation}
This model is exactly solvable by Bethe ansatz~\cite{Lieb63} and can be described by Luttinger liquid theory~\cite{Cazalilla03,Cazalilla04,Cazalilla11} in the low-energy limit. {By resorting to dimensional considerations, it has been shown that the LL model only depends on the dimensionless variable $\gamma \doteq g/ \rho_0$, which determines the weak ($\gamma \lesssim 1$) and strong ($\gamma \gg 1$) coupling limits~{\cite{Lieb63}}.} The Luttinger parameter as a function of such dimensionless coupling $\gamma$ has been shown to lie in the interval
\begin{equation}
K \in [1,\infty),
\label{eq:KLL}
\end{equation}
for all values of $\gamma$, resulting in a superfluid ground state.

The second example is the case of the screened Coulomb potential~\cite{Tsukamoto00,Inoue06}
\begin{equation}
w_{\rm C}(z) = \frac{C}{(z^ 2 + d^2) ^{\alpha + 1/2}},
\label{eq:wC}
\end{equation}
where $\alpha \ge 0$, $C$ defines the strength of the interaction and $d$ is a characteristic screening length~\cite{Tsukamoto00,Inoue06} or dimensional cutoff~\cite{Schulz93}. Genuine long-range Coulomb interaction is obtained by setting $\alpha = 0$ and $d\rightarrow 0$. By performing a bosonization analysis, it has been shown~\cite{Dalmonte10} that for $\alpha > 0$ and as a function of $C$ and $\mu$ the Luttinger parameter falls in the region
\begin{equation}
K \in (0,1].
\label{eq:KC}
\end{equation}
Moreover, it has been shown~\cite{Schulz93,Tsukamoto00} the existence of a crossover from a Luttinger liquid to a Wigner crystal as $\alpha \rightarrow 0$.

\begin{figure}
\centering
\includegraphics*[width=.9\columnwidth]{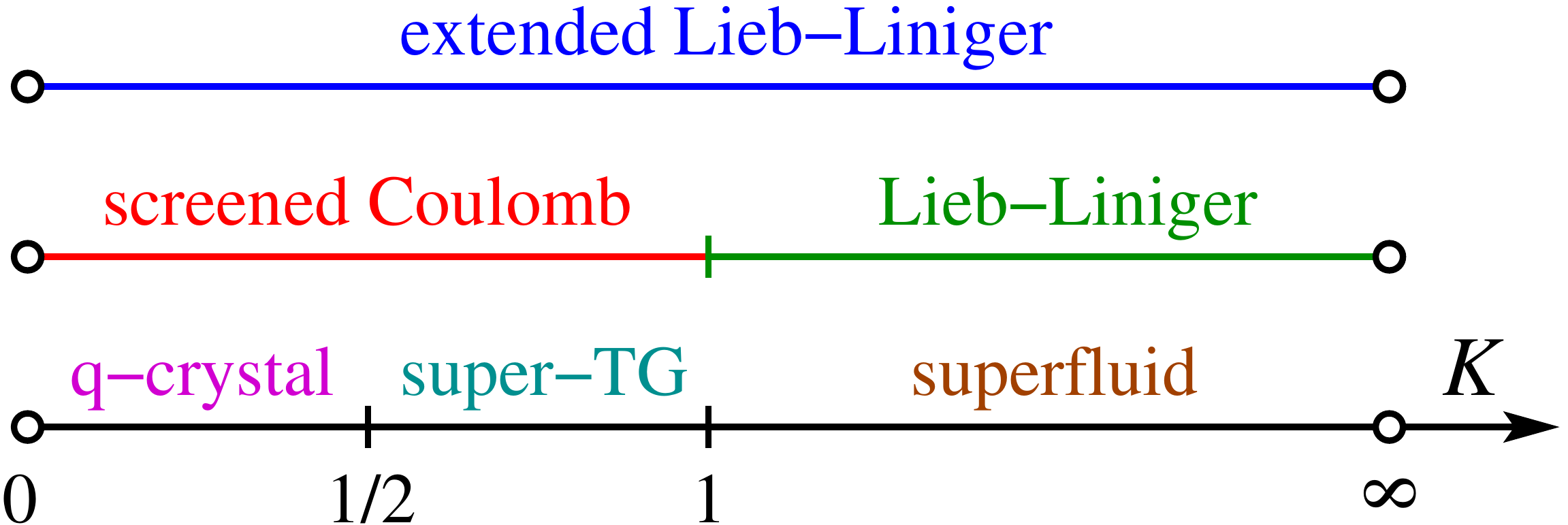}
\caption{(Color online) Values of the Luttinger parameter $K$ discriminated by the different interactions discussed in the text. For the LL model $K \in [1, \infty)$. In the case of screened Coulomb interactions $K \in (0, 1]$. And for the ELL model $K \in (0, \infty)$. The possible phases are quasi-crystal, super-Tonks-Girardeau, and superfluid. The Tonks-Girardeau gas corresponds to $K=1$.}
\label{fig:genK}
\end{figure}

Hamiltonian~(\ref{eq:realH}) corresponds to the choice of potential
\begin{equation}
w_{\exp}(x-y) = g\left(\frac{\eta}{2}e^{-\eta|x-y|}\right)
\label{eq:wexp}
\end{equation}
in Hamiltonian~(\ref{eq:H}). This Hamiltonian is completely specified by defining the parameter set $(g, \eta, \mu)$. In addition to the value of the effective interaction strength $\gamma$, potential~({\ref{eq:wexp}}) introduces the dimensionless parameter $\eta / \rho_0$, which controls its effective range (see Sec.~{\ref{sec:ana}}). In the spirit of lattice models of strongly correlated systems, we have dubbed this exponentially-decaying interacting system of bosons the \emph{extended Lieb-Liniger model}. We will show in this paper that by changing the values of $(g, \eta, \mu)$, we can obtain a Luttinger parameter that covers the interval
\begin{equation}
K \in (0, \infty).
\label{eq:Kexp}
\end{equation}
Therefore, the ELL model is a minimal model that contains the physics describe by both contact potential and screened long-range interactions. This is the main finding of our work. A pictorial representation of this quantity and its relation to Eqs.~(\ref{eq:wLL}) and (\ref{eq:wC}) is shown in Fig.~\ref{fig:genK}. In particular, we will see that the LL model results from our extended Hamiltonian on the limit $\eta \rightarrow \infty$, and that it also reproduces the long-distance physics of the long-range potential for finite $\eta$.

\section{Analytical Considerations\label{sec:ana}}
Let us start with a discussion on the physics of the exponentially-decaying interactions that will allow us to predict and interpret the numerical results presented in Sec.~\ref{sec:num}. We begin by noticing that the integrated strength of the potential~(\ref{eq:wexp}) gives
$
\int_{-\infty} ^{\infty} dz\, w_{\exp}(z) = g.
$ 
In particular, using the limit representation $\delta(z) = \lim_{\eta \rightarrow \infty} \frac{\eta}{2}\, e^{-\eta |z|}$ of the Dirac delta function, we see that indeed Hamiltonian~(\ref{eq:realH}) reduces to the LL model in the $\eta \rightarrow \infty$ limit,
\begin{equation}
w_{\rm LL}(z) = \lim_{\eta \rightarrow \infty} w_{\exp}(z).
\end{equation}

It is also enlightening to compare the potential $w_{\exp}(z)$ of Hamiltonian~(\ref{eq:realH}) to the screened Coulomb case shown in Eq.~(\ref{eq:wC}). For that, let us Fourier transform $w_{\exp}(z)$,
\begin{equation}
w_{\exp}(q) = \frac{1}{\sqrt{2\pi}}\, \frac{g}{1 + (q/\eta) ^ 2},
\label{eq:wofq}
\end{equation}
where $q$ stands for the momentum. We can then see that a quadratic term $w_{\exp} / g \sim 1 - (q/\eta)^2$ will be the leading order in the long-wavelength limit $(q/\eta \ll 1)$.

On the other hand, a Fourier analysis shows that the screened Coulomb potential can also be written as a quadratic function in the long-distance limit provided that $\alpha \ge 1$~\cite{Tsukamoto00,Inoue06}. To leading order in $q$, the interaction potentials of Eqs.~(\ref{eq:wC}) and~(\ref{eq:wexp}) are related through
%
$g \sim C/d ^{2\alpha}$ and $\eta \sim 1/d$.
%
Then both potentials reproduce the same leading behavior at long distance for $\alpha > 1$. In addition, we notice that exponentially-decaying interactions cannot stabilize a Wigner crystal. This can be seen by noticing that Eq.~(\ref{eq:wofq}) cannot produce the leading order $\log q$, which is a term necessary to stabilize such a state~\cite{Schulz93,Meyer09}. We notice as well that numerical calculations have shown that exponential potentials can mimic the general behavior of Eq.~(\ref{eq:wC})~\cite{Baker15}.

To explore some further properties of Hamiltonian~(\ref{eq:realH}), we introduce the scaling transformation: $x \rightarrow \tilde x = \lambda x$, {for the space coordinate $x$. This change of scale will induce a transformation on the field operators, as well, if we want to maintain canonical commutation relations: $[\psi^\dagger(\tilde x),\psi(\tilde y)] = \delta(\tilde x-\tilde y)$. Explicitly, this transformation is $\psi(x) \rightarrow \tilde\psi(x) = \lambda^{-\Delta}\psi(\lambda x)$, with $\Delta = -1/2$. We call $\Delta$ the canonical scaling dimension of the field $\psi(x)$. Next, we require that Hamiltonian~({\ref{eq:realH}}) remains invariant under the scaling transformation. In other words, if $H$ depends on the field $\psi$ and a set of coupling constants labeled by $\alpha$, then $H_{\alpha}[\psi(\tilde x)] = H_{\tilde\alpha}[\tilde\psi(x)]$, where $\tilde\alpha$ is a new set of couplings that in general depend on $\lambda$.}

{Applying the scaling transformation to Eq.~({\ref{eq:realH}}) we obtain a new transformed Hamiltonian, which we write down as}
\begin{widetext}
\begin{equation}
\lambda^2 H = 
\frac{1}{2m_0} \int dx\, \partial_x\psi^\dagger(x) \partial_x\psi(x)  - \mu(\lambda) \int dx\, \psi^\dagger(x)\psi(x) 
+ \frac{g(\lambda)}{2} \int dx \int dy \left(\frac{\eta(\lambda)}{2}\, e^{-\eta(\lambda) |x-y|}\right) \psi^\dagger(x)\psi^\dagger(y)\psi(y)\psi(x),
\end{equation}
\end{widetext}
where we have defined the rescaled couplings as
\begin{equation}
g(\lambda) = g_0 \lambda,\quad
\eta(\lambda) = \eta_0 \lambda,\quad
\mu(\lambda) = \mu_0 \lambda^2,
\end{equation}
and have redefined the `bare' parameters by attaching a subscript to them. We have conveniently written down $H$ as $\lambda^2 H$ so we can fix $m_0$ without changing the ground state wave function. Notice that if we choose $\lambda = \rho_0^{-1}$ then the dimensionless quantity $\eta / \rho_0$ defines the effective range of the interaction potential~({\ref{eq:wexp}}). From these scaling equations we can deduce that for any given $H_0$, defined by the parameter set $(g_0, \eta_0, \mu_0)$, we can obtain a Hamiltonian $H_{\lambda}$, defined by $(g(\lambda) , \eta(\lambda) , \mu(\lambda))$, by rescaling the couplings as prescribed by the relations above. In particular, this implies that if we calculate the correlation functions from the ground state of $H_0$ we can obtain those for the mapped $H_\lambda$ by rescaling the coordinates of the original correlations~\cite{DiFrancescoBook}. The scaling transformation also shows that Hamiltonian~(\ref{eq:realH}) has actually not three but two independent parameters, since they can be related by the scaling factor $\lambda$. For instance, we can eliminate $\eta$'s equation and write down
%
\begin{equation}
g(\lambda) = \frac{g_0}{\eta_0}\,\eta(\lambda), \quad \mu(\lambda) = \frac{\mu_0}{\eta_0^2}\,\eta(\lambda)^2.
\end{equation}

\begin{figure}
\centering
\includegraphics*[width=.85\columnwidth]{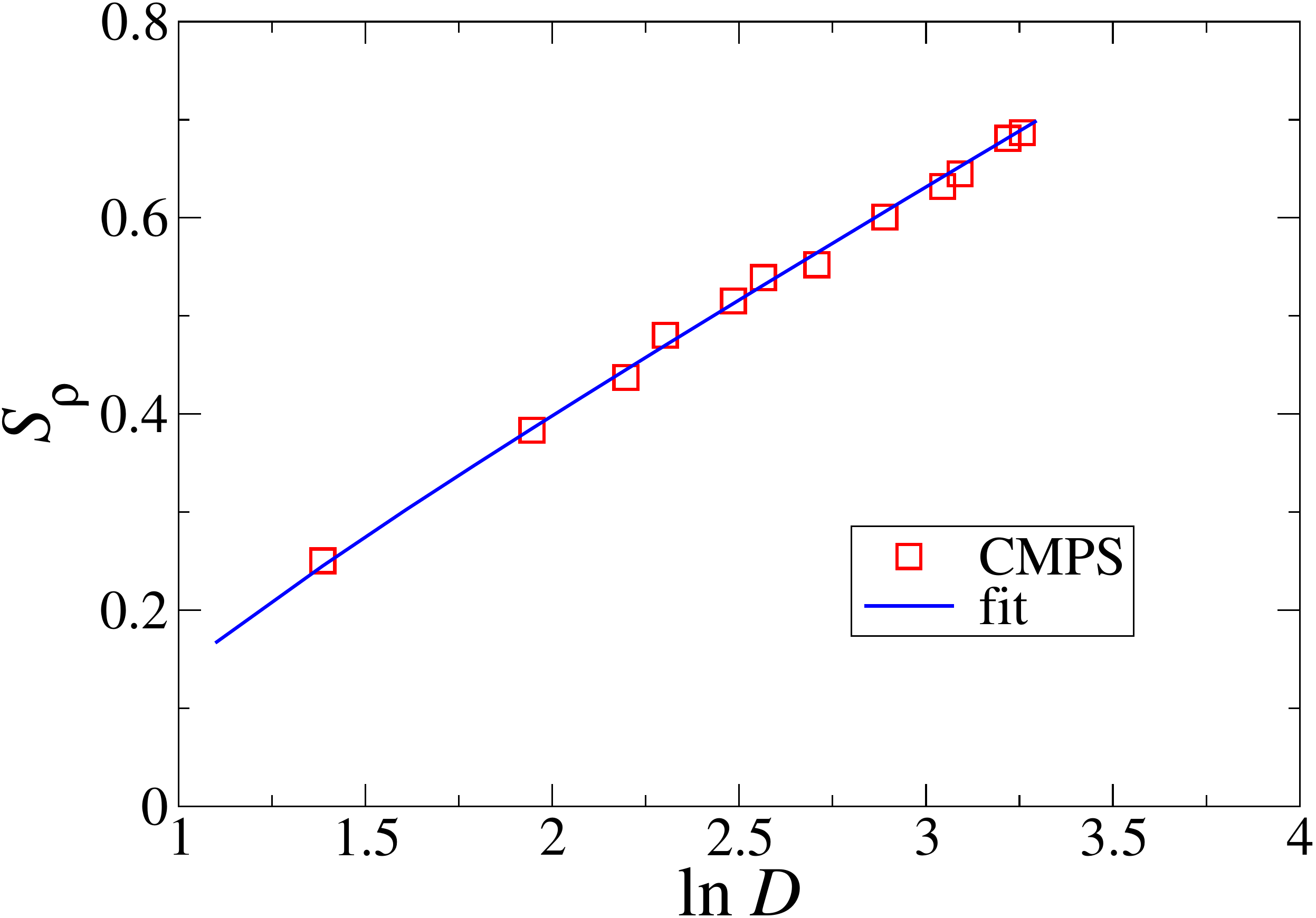}
\caption{(Color online) Entanglement entropy as a function of the bond dimension $D$ for $\gamma = 0.83$ and $\eta / \rho_0 = 3.29$. We have fitted the data to the scaling relation $S_\rho = \zeta \ln D + \mathcal{O}(1/\ln D)$, where $\zeta = (\sqrt{12/c} + 1)^{-1}$ and $c$ is the central charge~{\cite{Tagliacozzo08}}.}
\label{fig:Sent}
\end{figure}

As discussed at the beginning of this Section, the LL Hamiltonian is contained as a limiting case of the ELL model. This limit corresponds to $\eta(\lambda) \rightarrow \infty$ which in turn implies for the bare couplings that $\eta_0 \rightarrow \infty$, or $g_0 \rightarrow 0$ and $\mu_0 \rightarrow 0$, such that $g(\lambda)$ and $\mu(\lambda)$ remain finite. In the space of parameters defined by $(g, \eta, \mu)$ the LL case is constrained to the plane $\eta(\lambda) = \infty$ where $g(\lambda)$ and $\mu(\lambda)$ can take arbitrary values. As we shall see below, our numerical results indicate that for large, albeit finite, values of $\eta(\lambda)$ the ground state of Hamiltonian~(\ref{eq:realH}) behaves in a similar way to that of the LL model, for the same values of $(g(\lambda), \mu(\lambda))$. In the opposite limit, for values of $\eta(\lambda) \sim 1$, the physics of the ELL model is \emph{not} related to that of the standard LL Hamiltonian.

Similar conclusions on the scaling properties of Hamiltonian~({\ref{eq:realH}}) can be drawn by exploring the scaling of the dimensionless couplings $\gamma = g_0 / \rho_0$, $\mu_0 / \rho_0^2$, and $\eta_0 / \rho_0$. By introducing the transformation under change of scale of the density $\rho(\lambda) = \rho_0 / \lambda$, the rescaled couplings in this case transform as $\gamma(\lambda) = \gamma_0 \lambda^2$, $\eta(\lambda) / \rho(\lambda) = (\eta_0 / \rho_0) \lambda^2$, and $\mu(\lambda) / \rho(\lambda)^2 = (\mu_0 / \rho_0^2) \lambda^4$. The resulting set of independent equations now read
\begin{equation}
\gamma(\lambda) = \frac{\gamma_0}{\eta_0 / \rho_0} \left(\frac{\eta(\lambda)}{\rho(\lambda)}\right), \quad \frac{\mu(\lambda)}{\rho(\lambda)^2} = \frac{\mu_0}{\eta_0^2} \left(\frac{\eta(\lambda)}{\rho(\lambda)}\right)^2.
\end{equation}
The ELL model now defined by the couplings $(\gamma, \eta / \rho, \mu / \rho^2)$ will correspond to the LL model when $\eta(\lambda) / \rho(\lambda) \rightarrow \infty$. And as before, for $\eta(\lambda) / \rho(\lambda) \sim 1$, the ELL Hamiltonian will give rise to different phenomena than that of the LL model.

\section{Numerical Results\label{sec:num}}
We now discuss the numerical results obtained for the ELL model using the CMPS by Verstraete and Cirac~\cite{Verstraete10,Jutho13}, along with the time-dependent variational principle proposed by Haegeman~\emph{et al.}~\cite{Jutho10,Jutho11,Jutho13} as minimization method. The CMPS method produces an approximation to the ground state wave function. From this approximation we can compute quantities such as the ground state energy and particle densities as well as correlation~\cite{Verstraete10,Jutho10,Jutho13} and spectral~\cite{Draxler13} functions. For our translationallly invariant system, this variational ansatz is parametrized by two $D\times D$ matrices $Q$ and $R$ (see Appendix for details). These matrices contain the variational parameters of the ground state wave function. The bond dimension $D$ is a refining parameter that permits to control the accuracy of the resulting wave function (including expectation values and correlation functions). The computational cost grows as $\mathcal O(D^3)$.

{We have studied ground states for values of the bond dimension in the range $D = 4-32$. As a first check of the accuracy of the CMPS wave function in describing the ELL model, we have calculated the entanglement entropy in order to extract the value of the central charge. A typical result for the entanglement entropy as a function of $D$ is shown in Fig.~{\ref{fig:Sent}}. Using the finite-scaling entanglement formula for the entanglement entropy, proposed in Ref.~{\onlinecite{Tagliacozzo08}}, it is possible to extract an estimated value of the central charge of the conformal field theory underlying the ELL model. Such theory corresponds to the Luttinger liquid which possesses a central charge $c = 1$. The extracted value from the numerics $c \approx 0.95$ gives rise to an error of $5\%$ and compares fairly well with the expected result. We interpret this result as a confirmation that the CMPS gives an accurate approximation to the ground state wave function of the ELL model. In particular, this statement implies that the CMPS is capable of describing critical theories such as the Luttinger liquid, which is the low-energy effective theory of the ELL model (see Sec.~{\ref{sec:model}}).}

\begin{figure}
\centering
\includegraphics*[width=.85\columnwidth]{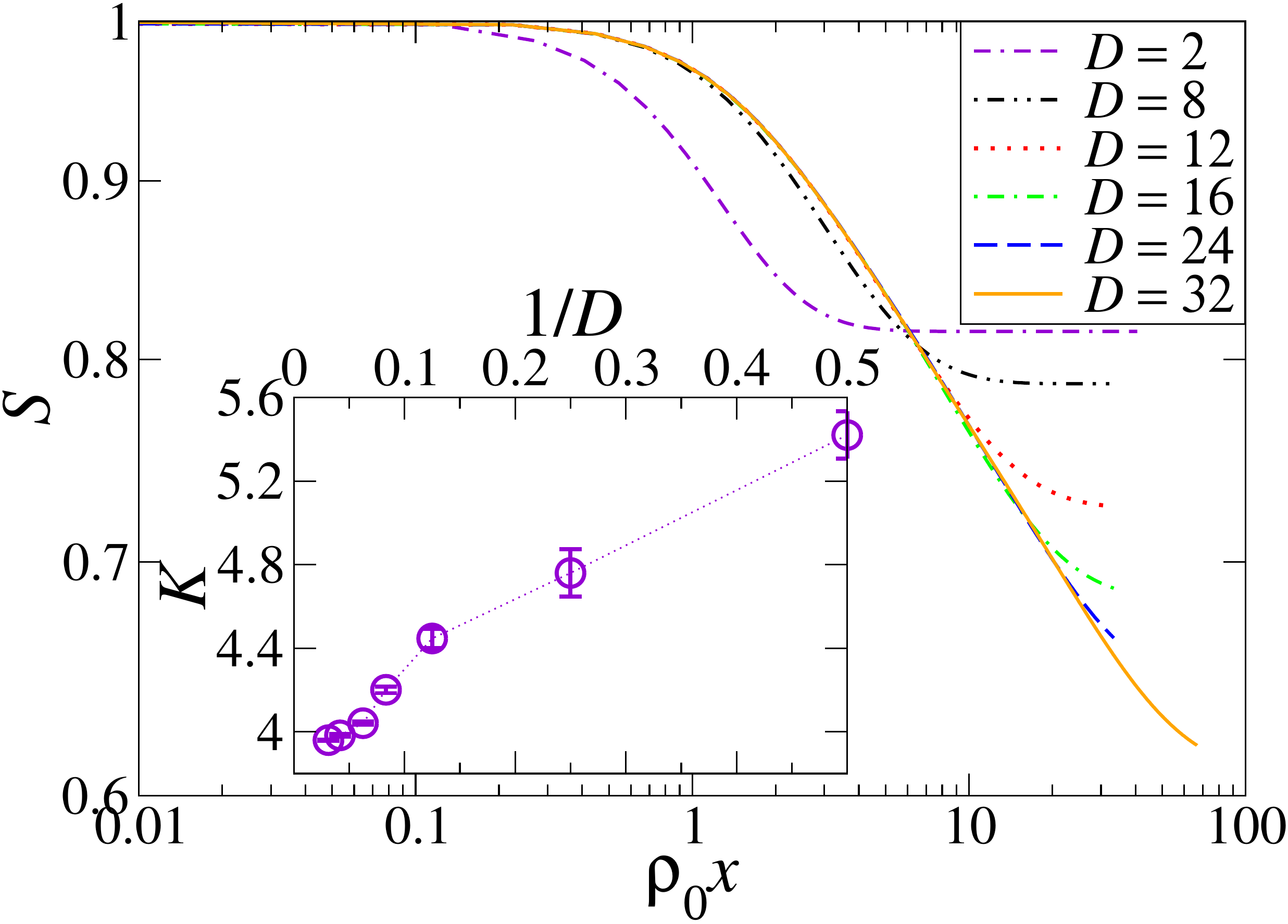}
\caption{(Color online) Superfluid correlation function as a function of distance for the LL model ($\eta \rightarrow \infty$) with {$\gamma = 0.75$}, and several values of $D$. Inset: Corresponding finite-entanglement scaling of the extracted $K$.}
\label{fig:fD}
\end{figure}

Within {the range of values of $D$ studied, }we have observed convergence of the quantities calculated in this paper, including the Luttinger parameter $K$. Figure~\ref{fig:fD} shows an example that correlation functions converge up to distances around 50 times the interparticle spacing, $x\sim 50\rho_0$. This plot also exhibits an increasingly large power-law region as $D$ increases. Likewise, $K$ versus $D$ shows a systematic convergence. $K$ has been extracted by fitting our numerical results of the correlation functions to Eq.~(\ref{eq:corrs}). The fitting variables are $K$, $A_1$, $B_0$, and $B_1$~\cite{NoteCmps}. {The extrapolation of $K(D)$ to $D \rightarrow \infty$ produces an estimate of the error of $K$ for a given $D$. In the example of Fig.~{\ref{fig:fD}}, assuming that $K(D)$ is quadratic in $1/D$, this error is $5\%$ for $D = 24$. Similar errors are obtained in the rest of results discussed in this paper. These results have been calculated with $D = 24$ unless otherwise stated.}

In the following, we will see that exponentially-decaying interactions lead to Luttinger parameters in the range of Eq.~(\ref{eq:Kexp}), thus containing the cases of screened Coulomb [Eq.~(\ref{eq:KC})] and contact [Eq.~(\ref{eq:KLL})] interactions. Having $K$ spanning such range will lead to crossovers from superfluid to super-Tonks-Girardeau to {quasi-crystal} states. Here we will refer to the super-Tonks-Girardeau regime as a state of suppressed superfluidity which can be described as spinless fermions (the Tonks-Girardeau gas) interacting repulsively~\cite{Cazalilla03,Astrakharchik08}.

\begin{figure}
\centering
\includegraphics*[width=.85\columnwidth]{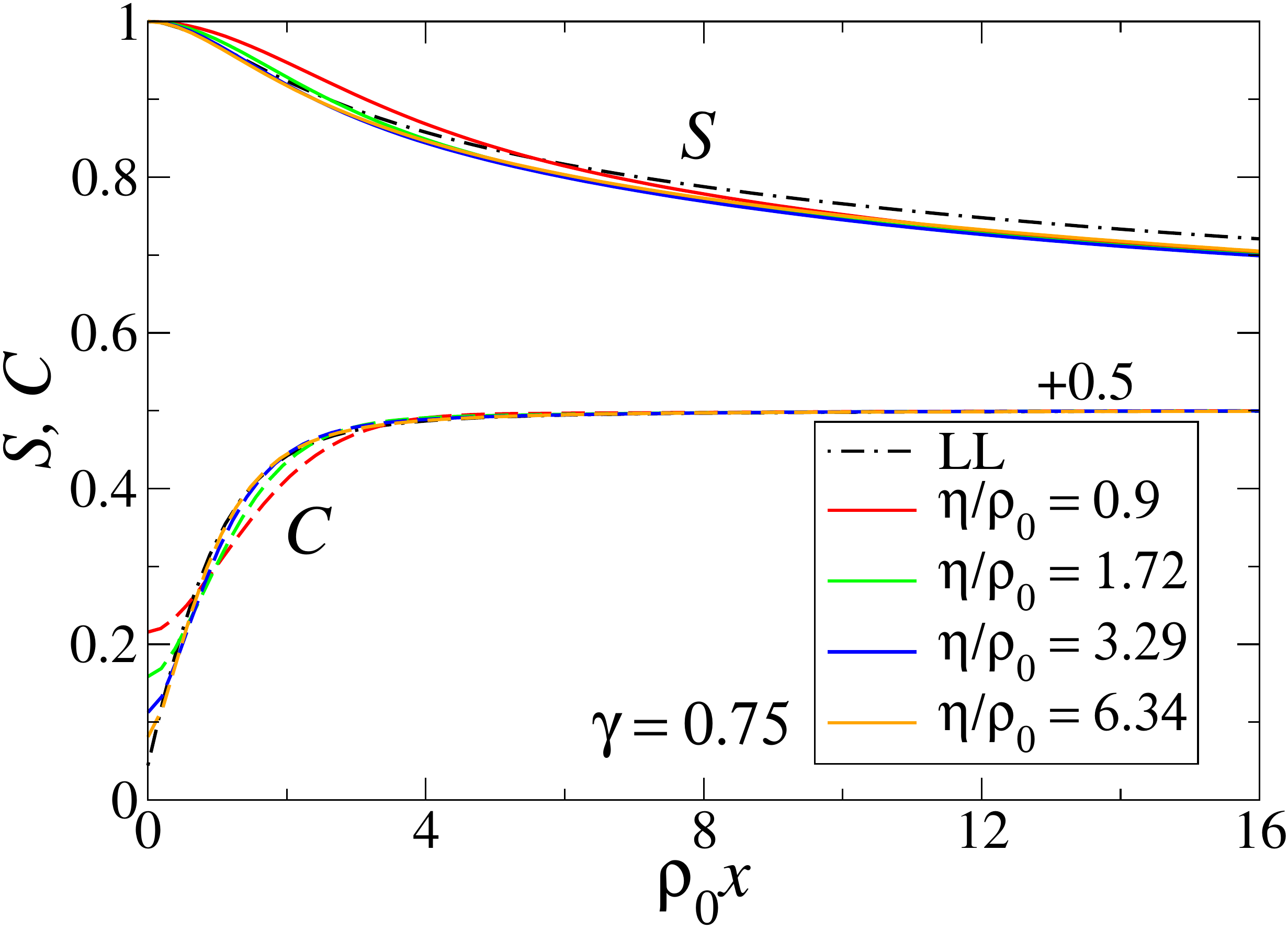}
\caption{(Color online) Weak coupling limit: {$\gamma = 0.75$}. Superfluid $S=S(\rho_0 x)$ (full) and density $C=C(\rho_0 x)$ (dashed) correlations as a function of the distance for several values of $\eta / \rho_0$. The dash-dotted lines are results for the LL case, see Eq.~(\ref{eq:wLL}). The continuous and dashed lines correspond to the ELL Hamiltonian~(\ref{eq:realH}). A superfluid ground state is found in this regime.}
\label{fig:wc}
\end{figure}

We will of focus on two relevant regimes: weak coupling, where $\gamma \lesssim 1$, and the strong coupling limit, for which $\gamma \gg 1$. We have fixed the chemical potential to $\mu = 0.5$ throughout. Results for other {values of the couplings} are connected via a scaling transformation, as discussed in Sec.~\ref{sec:ana}. For each limiting case, we have varied the value of the characteristic length of the potential $\eta$ and follow the evolution of the correlation functions.

\subsection{Weak Coupling\label{sec:wc}}
Let us start by analyzing the weak coupling limit for {$\gamma = 0.75$}. Figure~\ref{fig:wc} shows the results for the density $C(x)$ and superfluid $S(x)$ correlation functions as a function of the distance scaled to the density $\rho_0$. {As can be seen, the superfluid correlations closely resemble those of the the LL case (shown as the dash-dotted black line). This suggests that the values of the Luttinger parameter of the extended model will be presumably similar to the LL model, and only differences will appear at short distances, $\rho_0 x \lesssim 1$. The density correlations confirm these findings. Indeed, for $\rho_0 x \gg 1$ the long distance behavior of the extended and standard LL models is closely related. Again differences are found at short distances $\rho_0 x \sim 1$, \emph{i.e.} at energies not reachable with the theory of Luttinger liquid.}

\begin{figure}
\centering
\includegraphics*[width=.85\columnwidth]{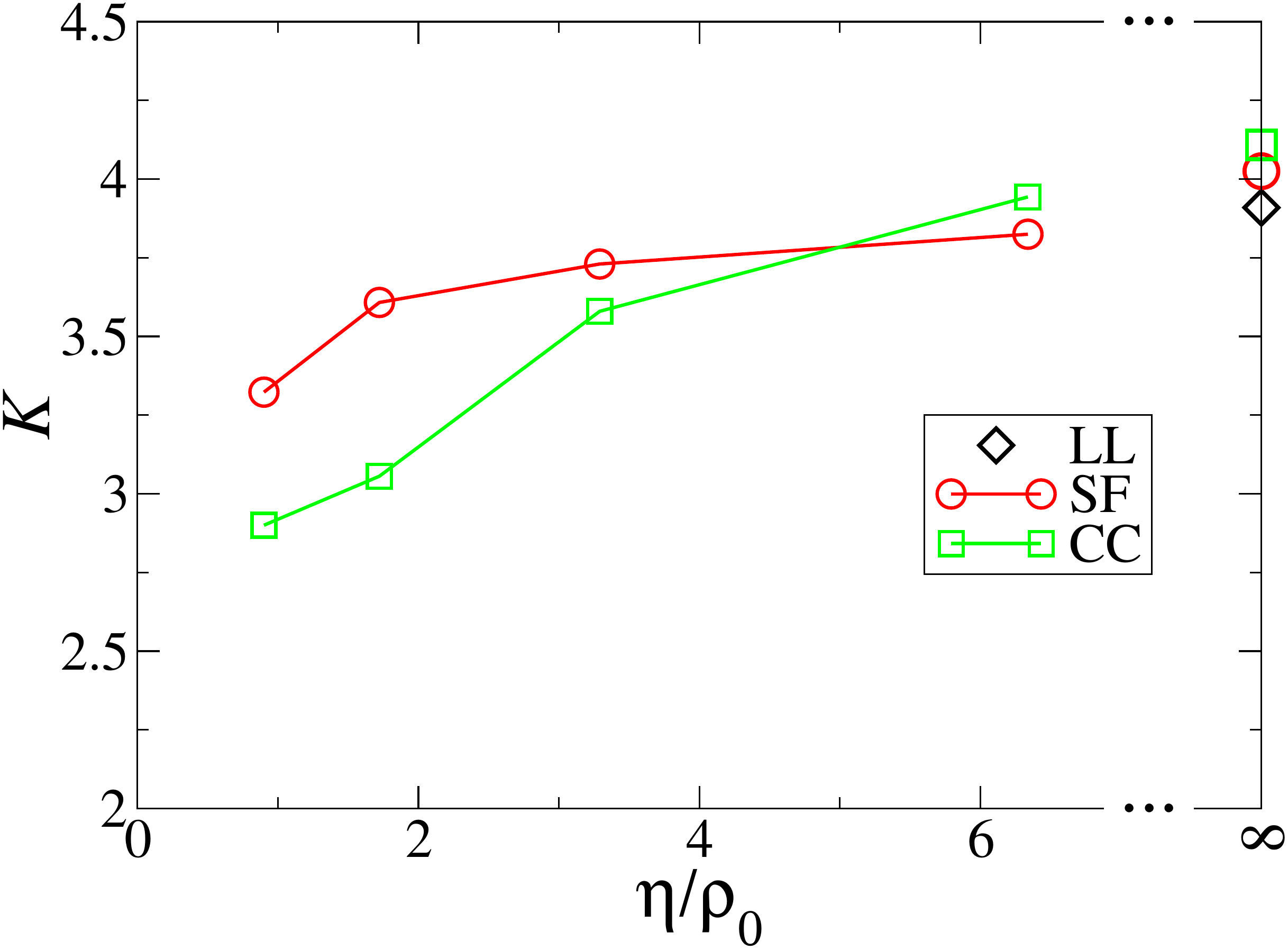}
\caption{(Color online) Luttinger parameter $K$ as a function of the potential effective range $\eta / \rho_0$ at weak coupling. The values of $K$ extracted from the fittings to the data in Fig.~\ref{fig:wc} for the superfluid (charge density) correlations are labelled as SF (CC). For the LL model $(\eta \rightarrow \infty)$ the label LL corresponds to the values of $K$ extracted from bosonization formulae~(\ref{eq:corrs})~\cite{Cazalilla04}.}
\label{fig:Kwc}
\end{figure}

The resulting values of $K$ at weak coupling for the data of Fig.~\ref{fig:wc} are shown in Fig.~\ref{fig:Kwc}. {Our results for the LL case ($\eta \rightarrow \infty$) compare fairly well to the values of $K$ predicted using bosonization, giving an error of $\sim 5\%$. As was discussed for Fig.~{\ref{fig:wc}}, we observe an overall superfluid state in the ELL Hamiltonian at weak coupling; meaning $K > 1$ according to Eq.~({\ref{eq:corrs}}). Figure~{\ref{fig:Kwc}} shows that $K > 1$ for the ELL Hamiltonian, leading to $S(x)$ decaying the slowest and hence to a superfluid ground state. As $\eta / \rho_0$ grows, the value of $K$ is closely related to that of the LL model, as expected.}

It is possible to show what the possible values of $K$ are for the ELL model in the weak coupling limit, see Eq.~(\ref{eq:Kexp}). Starting at $\eta \rightarrow \infty$, we know that the exponentially-decaying interactions~(\ref{eq:wexp}) contain the LL model. On the other hand, for finite but large $\eta / \rho_0$ we have observed as well superfluid behavior ($K > 1$), and as $\eta / \rho_0$ decreases the prefactor $g \eta$ in Hamiltonian~(\ref{eq:realH}) will be small, leading to quasi-free bosons, which imply $K \rightarrow \infty$. Consequently we can expect that, at least in the weak coupling limit, the ELL model will have a Luttinger parameter in the region $K \in [1,\infty)$. Notice that this interval is the same for $K$ in the LL model show in Eq.~(\ref{eq:KLL}), see also Fig.~\ref{fig:genK}.

\subsection{Strong Coupling\label{sec:sc}}
\begin{figure}
\centering
\includegraphics*[width=.85\columnwidth]{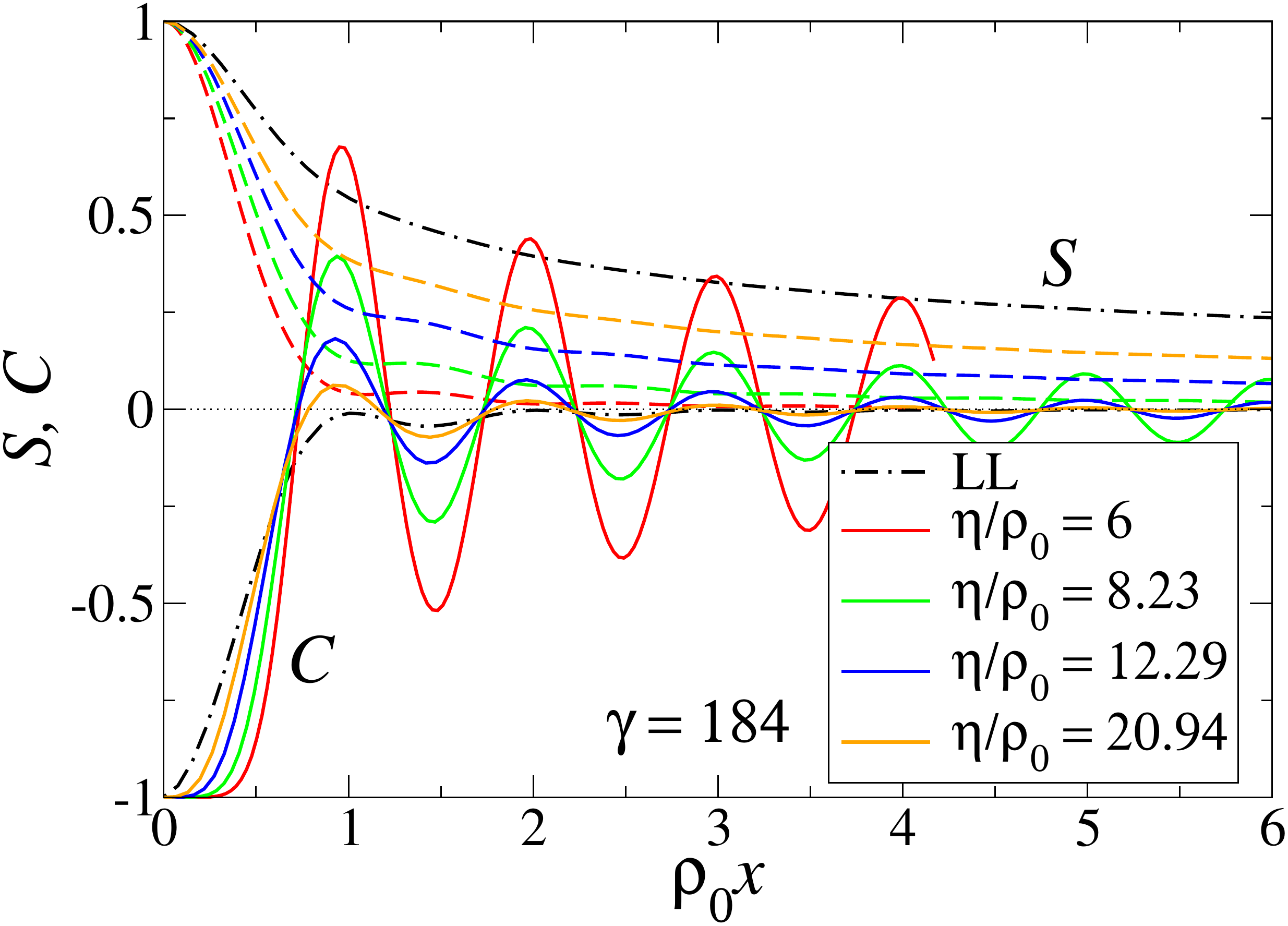}
\caption{(Color online) Strong coupling limit: {$\gamma = 184$}. The notation is the same as in Fig.~\ref{fig:wc}. In this limit the superfluid correlations are strongly suppressed whereas the charge correlation function displays an increase indicating a crossover from the super-Tonks-Girardeau to a quasi-crystal state.}
\label{fig:sc}
\end{figure}

The strong coupling limit results of the correlations are shown in Fig.~\ref{fig:sc}. Firstly, for the LL case where $\eta \rightarrow \infty$ (dash-dotted line), the dominant correlations are those of a superfluid ground state, similarly to weak coupling. The only remarkable difference is that for $\gamma \gg 1$ the charge correlations display Friedel oscillations characteristic of the Tonks-Girardeau regime~\cite{Girardeau60,Arkhipov05}, where the bosonic system maps to free spinless fermions~\cite{Cazalilla03,Astrakharchik08}. Secondly, for the extended model at large $\eta / \rho_0$, superfluidity is suppressed although still remains as the dominant fluctuation at long distances. However, charge correlations become increasingly large in the short distance. Notice that for large but finite $\eta / \rho_0$, superfluidity is greatly suppressed compared to the LL model. An overall increase of Friedel oscillations is also observed in the charge sector.

Further decreasing of $\eta / \rho_0$ leads to an almost complete suppression of superfluid correlations and charge fluctuations thus govern at short and long distances. In the range where charge correlations decay slower than superfluid ones, we observe the appearance of a definite wavevector $Q$ that modulates the density fluctuations. As expected from the correlation functions~(\ref{eq:corrs}), this wavevector is set by $\rho_0$. The average density sets a length scale $a = 1 / \rho_0$ and we can associate a wavevector to it as $Q = 2 \pi / a = 2 \pi \rho_0$. The appearance of this wavevector in $C(x)$ signals the establishment of {charge order}. 

\begin{figure}
\centering
\includegraphics*[width=.85\columnwidth]{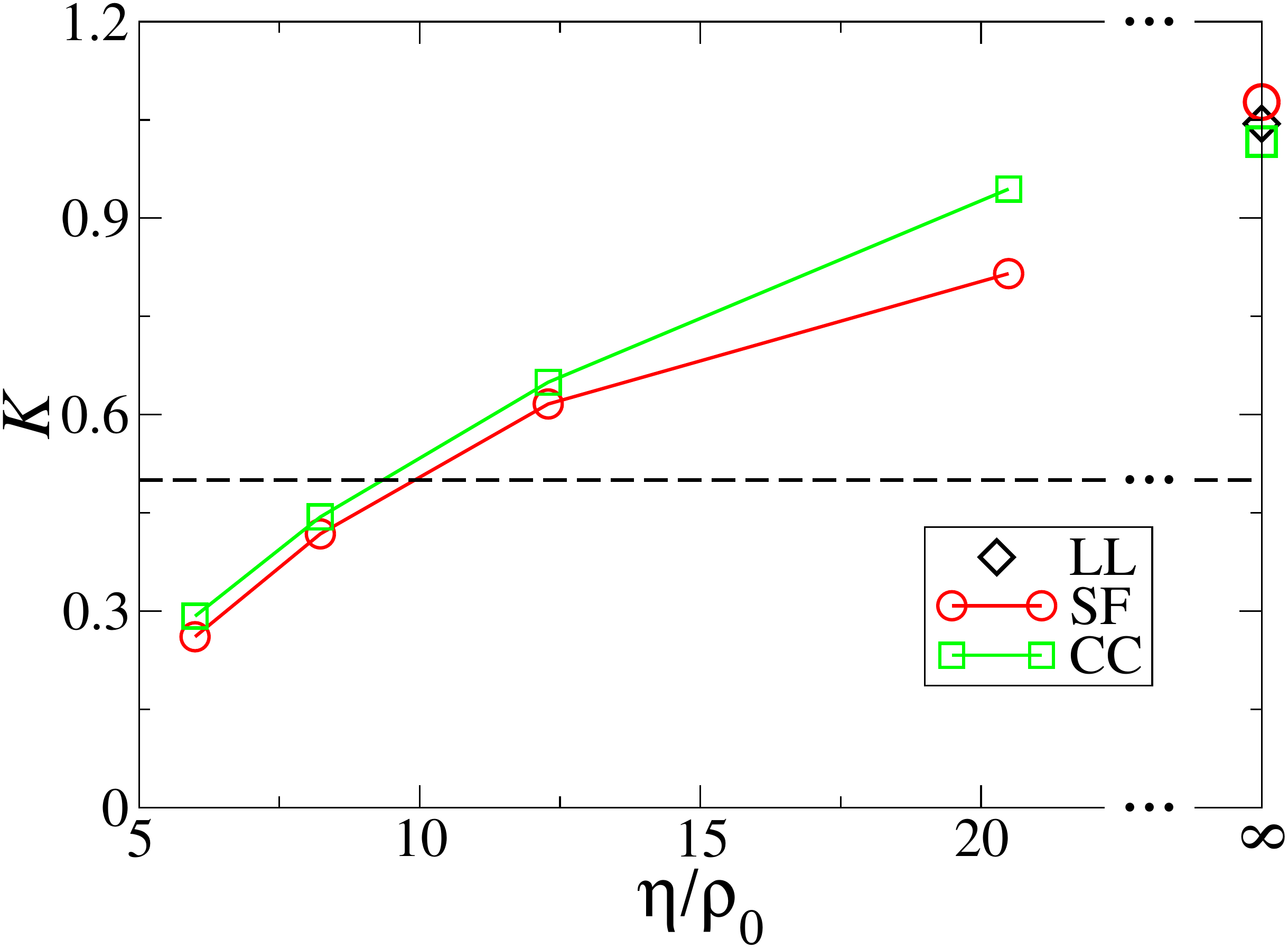}
\caption{(Color online) Luttinger parameter $K$ as a function of the potential effective range $\eta / \rho_0$ at strong coupling. The values of $K$ extracted from the fittings to the data in Fig.~\ref{fig:sc}. The rest of the notation is the same as in Fig.~\ref{fig:Kwc}. The crossover value at $K = 1/2$ from the super-Tonks-Girardeau limit to the quasi-crystal state is shown as a dashed line. See text for details.}
\label{fig:Ksc}
\end{figure}

Figure~\ref{fig:Ksc} displays the results of the Luttinger parameter at strong coupling. As can be seen, the value of $K$ is restricted to $K < 1$, which indicates that the ensuing physics is \emph{not} related to the LL model. As $\eta / \rho_0$ in varied, $K$ increases and presumably when $\eta \rightarrow \infty$ the results of the LL model are recovered. The estimated error of $K$ for the LL model is around $3\%$. By inspecting Eq.~(\ref{eq:corrs}) a crossover from a state with suppressed superfluidity to a quasi-crystal state is obtained when $K = 1/2$. Accordingly, we expect that upon decreasing $\eta / \rho_0$ (\emph{i.e.},~away from the LL limit) the tail of the exponential interaction will dominate, leading to the formation of the ordered state for $K < 1/2$. For $K > 1/2$ the bosonic system is described by strongly interacting repulsive spinless fermions; this is a suppressed superfluid state or the super-Tonks-Girardeau regime.

The overall picture is then that, for a fixed value of {$\gamma$} and by varying $\eta / \rho_0$, a crossover from a {super-Tonks-Girardeau to a charge-ordered} state is observed. This ordered state is in fact a {quasi-crystal}, which is the closest state that resembles a Wigner crystal within Luttinger liquid theory. The presence of the super-Tonks-Girardeau and {quasi-crystal} states, which are not present in the LL Hamiltonian, is directly related to the exponentially-decaying potential~(\ref{eq:wexp}).

We have performed similar calculations for larger values of {$\gamma$} (not shown here). The general tendency is similar to that shown in Fig.~\ref{fig:sc}. However, the range of values of $K$ is even smaller than the values reported in Fig.~\ref{fig:Ksc}. This suggests that in the strong coupling limit of the ELL Hamiltonian, the Luttinger parameter can range in the interval $K \in (0, 1]$. This assumption is supported by the discussion of Sec.~\ref{sec:ana}. There we have shown that Hamiltonian~(\ref{eq:realH}), with specific values of $(g, \eta)$, can describe screened Coulomb interactions in the long-wavelength limit (see Fig.~\ref{fig:genK}). In addition, bosonization results on systems interacting through potential~(\ref{eq:wC}) have shown that the compactification radius lies in the same region as $K$ in the ELL model for $\gamma \gg 1$. Notice that such values of $K$ at strong coupling match those of Eq.~(\ref{eq:KC}).

\subsection{Luttinger Parameter\label{sec:lp}}
\begin{figure}
\centering
\includegraphics*[width=.85\columnwidth]{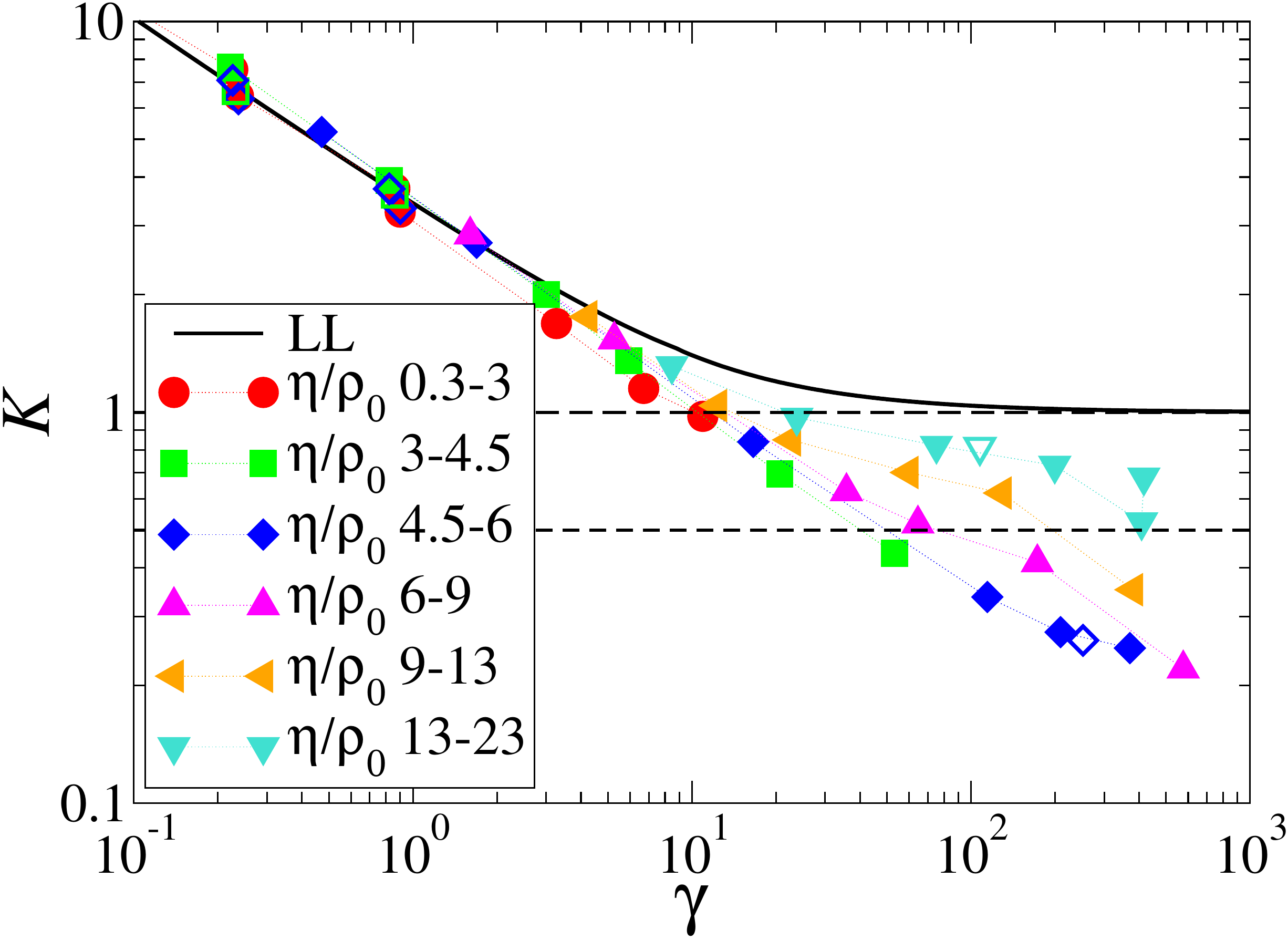}
\caption{(Color online) Luttinger parameter $K$ versus {the dimensionless interaction $\gamma = g / \rho_0$} for color-coded intervals of $\eta / \rho_0$, for the ELL model. Full (open) symbols correspond to $D = 18~(24)$. The values of the effective range of the potential span $\eta / \rho_0 \in [0.3, 23]$. The full line is the bosonization result for the standard LL model, obtained with Luttinger liquid theory~\cite{Cazalilla04}. The dashed lines correspond to the superfluid/super-Tonks-Girardeau and super-Tonks-Girardeau/quasi-crystal crossovers at $K = 1$ and $K = 1/2$, respectively.}
\label{fig:KvsG}
\end{figure}

The analysis and conclusions drawn in Subs.~\ref{sec:wc} and~\ref{sec:sc} can be further substantiated by calculating the Luttinger parameter for arbitrary values of $(g, \eta)$, with $\mu = 0.5$. Results for other {coupling values} are connected via the scaling transformation discussed in Sec.~\ref{sec:ana}. In Fig.~\ref{fig:KvsG} we show the Luttinger parameter, for the ELL model, versus $\gamma$ for several color-coded intervals of $\eta / \rho_0$. The data confirms our expectations that $K \in (0, \infty)$, showing that Hamiltonian~(\ref{eq:realH}) contains as limiting cases the LL model and the screened Coulomb potential.

As discussed for the weak coupling ($\gamma \ll 1$) case in Subs.~\ref{sec:wc}, the behavior of $K$ for the ELL model matches that of the weak-coupling regime of the standard LL (full line in Fig.~\ref{fig:KvsG}) model. This indicates that the effect of the potential range does not greatly affect the physics of the extended model for $\gamma \ll 1$. In this regime, superfluid correlations govern the ground state, \emph{i.e.}, $K > 1$.

At intermediate coupling, $\gamma \gtrsim 1$, the exponentially-decaying potential starts changing the level of correlations, and a departure from the LL result is observed, depending on the value of $\eta / \rho_0$. Indeed for large $\eta / \rho_0$ a LL-like trend is still seen, in accord with the results discussed in Subs.~\ref{sec:wc}. On the other hand, for smaller $\eta / \rho_0$ the Luttinger parameter crosses the Tonks-Girardeau point $K = 1$ into the super-Tonks-Girardeau state, where superfluidity is suppressed (see Subs.~\ref{sec:sc}).

Finally, for strong interactions, where $\gamma \gg 1$, a crossover from the super-Tonks-Girardeau regime to the quasi-crystal state is seen as a function of both fixed $\eta / \rho_0$ and increasing $\gamma$ and fixed $\gamma$ and decreasing $\eta / \rho_0$. The crossover line is defined by $K = 1/2$ (see Subs.~\ref{sec:sc}). Figure~\ref{fig:KvsG} shows that for large $\eta / \rho_0$ the strong coupling limit of the extended model tends to closely follow that of the LL model. As $\eta / \rho_0$ decreases an overall superfluid/super-Tonks-Girardeau/quasi-crystal crossover is observed. 

Based on the previous discussions, we conclude that for arbitrary values $(g, \eta, \mu)$ the low-energy physics of the ELL Hamiltonian is described by Luttinger liquid theory with a Luttinger parameter lying in the range~(\ref{eq:Kexp}). This result is in high contrast with the cases of contact [Eq.~(\ref{eq:KLL})] and of power-law [Eq.~(\ref{eq:KC})] interactions (see Fig.~\ref{fig:genK}). 
This entails that by tuning $(g, \eta, \mu)$ the decaying exponential can be short-ranged enough to describe LL physics or \emph{power-lawed} enough to obtain an analog behavior to that observed for the screened Coulomb interaction.

\section{Conclusions\label{sec:con}}
Let us summarize the main results presented in this paper. Firstly, we have introduced a model for bosons in $1 + 1$ dimensions interacting through an exponentially-decaying potential. Secondly, by employing well-established scaling transformations we have shown that this ELL model contains, in some limiting cases, both the standard LL model and the long-wavelength limit of the screened Coulomb potential. This discussion allowed us to make some predictions on the phases of exponentially-decaying interacting bosons such as the presence of {quasi-crystal} and super-Tonks-Girardeau states. Such states are not present in the original LL model, for which the ground state is always superfluid. Thirdly, making use of the recently developed CMPS techniques we have explored the ground state phase diagram of our system of exponentially-interacting bosons.

{By calculating the superfluid and density correlation functions, we have shown that at weak coupling superfluidity governs the ground state in much the same way as in the LL model.} At strong coupling, however, superfluidity is strongly suppressed with a simultaneous increase of density correlations, signaling the emergence of the super-Tonks-Girardeau state. Upon increasing the interaction density correlations dominate the ground state and a {quasi-crystal state} is stabilized. As noticed above, these additional phases are not present in the original LL Hamiltonian. Hence, the decaying exponential potential induces a crossover from superfluid to super-Tonks-Girardeau to {quasi-crystal} states, when the interaction strength is varied from weak to strong coupling. Finally, resorting to Luttinger liquid theory, we have shown that the value of the Luttinger parameter ranges in the interval $K \in (0, \infty)$; thus differing from the values taken by the corresponding quantity of both the LL model, where $K \in (1, \infty)$, and the screened Coulomb case, for which $K \in (0, 1)$.

\begin{acknowledgments}
The authors thank B. Paredes for insightful discussions. J. R.~acknowledges fruitful conversations with J. Carrasquilla, T. Baker, M. Stoudenmire, and A. Schlief. M. G.~gratefully acknowledges interesting discussions with F. Verstraete and D. Draxler.
The authors also acknowledge support by the Simons Foundation (Many Electron Collaboration). This research was supported in part by Perimeter Institute for Theoretical Physics. Research at Perimeter Institute is supported by the Government of Canada through Industry Canada and by the Province of Ontario through the Ministry of Economic Development \& Innovation.
\end{acknowledgments}

\appendix*

\section{Uniform Continuous Matrix Product States for Exponential Interactions\label{app:cmps}}
In this paper we numerically investigate translationally invariant field theories of interacting bosons in $1+1$ dimensions, using the technique of continuous matrix product states (CMPS) of Verstraete and Cirac~\cite{Verstraete10}. In this appendix we give a more detailed discussion of the formalism and its implementation.

Similar to the case of MPS~\cite{verstraete_matrix_2008,schollwock_density-matrix_2011} on a lattice, the method makes an ansatz for a ground state wave function $\ket{\Psi}$ in terms of a set of continuous, matrix-valued functions $Q(x)$ and $R(x)$ on an interval $x\in [-L/2,L/2]$:
\begin{equation}
\ket{\Psi}=v_l^{\dagger}\mathcal{P}e^{\int_{-L/2}^{L/2}dx\, Q(x)\otimes \mathbbm{1}+R(x)\otimes\psi^{\dagger}(x)}v_r|0\rangle.
\label{eq:cmps}
\end{equation}
Here, $Q(x)$ and $R(x)$ are $D\times D$ matrices for every point $x$ (comprising the variational space), $v_l$ and $v_r$ are boundary vectors at $x=\pm L/2$ which incorporate boundary conditions, $\mathcal{P}e$ is the path-ordered 
exponential, $\mathbbm{1}$ and $\psi^{\dagger}(x)$ are identity and creation operators acting at position $x$ in space, 
and $\ket{0}$ is the vacuum defined by $\psi(x)|0\rangle = 0$. $D$ is called the bond dimension of the CMPS. When going to the thermodynamic limit $L\rightarrow \infty$ (see below), $v_l$ and $v_r$ will drop out of any equations and can be neglected. For later reference we define 
\begin{equation}
U(x,y) = \mathcal{P} \exp{\int_x^y dx\,Q(x)\otimes\mathbbm{1} + R(x)\otimes\psi^{\dagger}(x)}.
\end{equation}

The CMPS \Eq{eq:cmps} can be considered to be the limit of a certain type of lattice MPS: consider a discretization of the interval $[-L/2,L/2]$ into $N$ equidistant points $x_n$, separated by $\epsilon$. One can show~\cite{Verstraete10} that when expanding the path-ordered exponential, defining $c_n^{\dagger}\equiv \sqrt{\epsilon}\psi^{\dagger}(x_n)$ and collecting orders of $\epsilon$, the resulting expression is equivalent to the one obtained from an MPS 
\begin{equation}
  \ket{\phi} = \sum_{\{i_n\}}
  A^{i_1}A^{i_2}\cdots A^{i_N}(c_1^{\dagger})^{i_1}(c_2^{\dagger})^{i_2}\cdots (c_N^{\dagger})^{i_N}|0\rangle
\end{equation}
with matrices $A^{i_n}$ restricted to the form
\begin{equation}
\begin{split}
  A^{i_n=0}&=\mathbbm{1}+\epsilon\, Q(x_n)\\
  A^{i_n=k>0}&=\frac{\sqrt{\epsilon^k}}{k!}R^k(x_n).
\end{split}
\end{equation}

Like for lattice MPS, the goal is to approximate the ground state wave function of a Hamiltonian $H=\int dx\, h(x)$ in terms of a CMPS. Time evolution is done using the time dependent variational principle, proposed by Haegeman \emph{et al.}, for CMPS~\cite{Jutho11}. In this method, time evolution is carried out by constructing $\frac{d}{d\tau}\ket{\Psi(\tau)}=-H|\Psi(\tau)\rangle$ and using it to update the wave function $\ket{\Psi}$:
\begin{equation}
  \ket{\Psi(\tau+d\tau)}=\ket{\Psi(\tau)}-d\tau H|\Psi\rangle.
\end{equation}
{The time dependent variational principle teaches us that the best optimal approximation to $H|\Psi\rangle$ is given by a tangent vector (see below) belonging to the tangent space to variational manifold of the CMPS. This implies a projection of $H|\Psi\rangle$ onto the tangent space, at which point the procedure becomes approximate (see Ref.~{\onlinecite{Jutho11}} for details).}

The goal is thus to find a {tangent vector} $\ket{\Phi}$ of given fixed bond dimension $D$ which optimally approximates $H|\Psi\rangle$, such that $\ket{\Psi}-d\tau|\Phi\rangle$ is again a CMPS of bond dimension $D$. Usually $H$ is a sum of local operators and thus a reasonable ansatz for $\ket{\Phi}$ is given by locally varying $Q(x)$ and $R(x)$, and taking a superposition of all these variations~\cite{Jutho11}:
\begin{multline}
  \ket{\Phi}=\int_{-L/2}^{L/2}dx\, U(-L/2,x)
  \\ \times (V(x)\otimes \mathbbm{1}+W(x)\otimes\psi^{\dagger}) U(x,L/2)|0\rangle.
\end{multline}
$V(x)$ and $W(x)$ are the variations of $Q(x)$ and $R(x)$, respectively. When added to $\ket{\Psi}$, the resulting state is again a CMPS of bond dimension $D$. Such vectors are also referred to as tangent vectors~\cite{Jutho11,Jutho13}. The optimal $V^*(x)$ and $W^*(x)$ are determined by minimization:
\begin{equation}
\{V^*, W^*\} = \text{argmin}_{\{V,W\}}\lVert\ket{\Phi}-H|\Psi\rangle\rVert^2
\label{eq:min}
\end{equation}
For translationally invariant systems in the thermodynamic limit ($L\rightarrow\infty$), which is what we consider in the following, $Q(x),\,R(x),\,W(x)$, and $V(x)$ can be chosen to be independent of $x$.

A gauge transformation~\cite{verstraete_matrix_2008,schollwock_density-matrix_2011,Jutho13} for a CMPS is a transformation on $(Q,R)$ which leaves $\ket{\Psi}$ invariant. It induces a redundancy in the tangent space, because it implies~\cite{Jutho11,Jutho13,haegeman_thesis} the existence of certain choices of non-zero $V_0$ and $W_0$ such that the resulting tangent vector $\ket{\Phi}$ to the state $\ket{\Psi}$ is zero: $\ket{\Phi[V_0,W_0]}\equiv 0$. Such undesirable variations of $\ket{\Psi}$ can be excluded by choosing a particular parametrization of $V$ and $W$ (other choices are possible~\cite{Jutho13}):
\begin{equation}
\begin{split}
V &= -l^{-1}R^{\dagger}l^{1/2}Yr^{-1/2},\\
W &= l^{-1/2}Y r^{-1/2},
\label{eq:tangentparam}
\end{split}
\end{equation}
$l$ and $r$ are the left and right reduced density matrices~\cite{Jutho13}, obtained from solving the equations
\begin{equation}
\begin{split}
\frac{d(l|}{dx} = (l|T &= (l|\left(Q\otimes \mathbbm{1}+\mathbbm{1}\otimes \bar Q+R\otimes \bar R\right)\\
& = lQ+Q^{\dagger}l+R^{\dagger}lR = 0\\
-\frac{d|r)}{dx} = T|r) &= \left(Q\otimes \mathbbm{1}+\mathbbm{1}\otimes \bar Q+R\otimes \bar R\right)|r)\\
& = Qr+rQ^{\dagger}+RrR^{\dagger} = 0.
\end{split}
\end{equation}
$T$ is called the transfer operator, {and acts as a superoperator on the vectors} $(l|$ and $|r)$. $l$ is in this respect a reordering of the vector $(l|$ into a matrix. We use the convention $(l|A\otimes\bar B\equiv B^{\dagger}lA$ and $A\otimes\bar B|r)\equiv ArB^{\dagger}$. To order $\epsilon$, the operator $e^{\epsilon T}$ equals the MPS transfer matrix $E=\sum_{i_n} A^{i_n}\otimes \bar A^{i_n}$~\cite{mcculloch_from}.

For our simulations we use the gauge freedom~\cite{Verstraete10} to fix the gauge of $\ket{\Psi}$ such that $l=\mathbbm{1}$. \Eq{eq:tangentparam} enforces $\braket{\Phi|\Psi}=0$; furthermore, we have
\begin{align}
\braket{\Phi|\Phi} = \delta(0)\,{\rm tr}(YY^{\dagger}).
\label{eq:ov}
\end{align}
We will consider the Hamiltonian~(\ref{eq:realH}).
Using pa\-ra\-me\-tri\-za\-tion~(\ref{eq:tangentparam}), $\langle\Phi|H|\Psi\rangle$ can be evaluated to~\cite{Draxler13,haegeman_thesis}
\begin{widetext}
\begin{multline}\\[-30pt]
  \langle\Phi|H|\Psi\rangle = \delta(0)\bigg[(l|\left(\frac{1}{2m}[Q,R]\otimes[\bar Q, \bar R] - \mu\, R\otimes \bar R
  + R\otimes\bar R\, \mathcal{L}[w](-T)R\otimes\bar R\right)
  (-T)_P^{-1}\left(\mathbbm{1}\otimes\bar V+R\otimes\bar W\right)|r)\\
  +(l|(R\otimes \bar R)\mathcal{L}[w](-T)\left(\mathbbm{1}\otimes\bar V+R\otimes\bar W\right)\mathcal{L}[w](-T)(R\otimes \bar R)|r)
  +(l|\Big(\frac{1}{2m}[Q,R]\otimes \left([\bar Q,\bar W]+[\bar V, \bar R]\right)\\ - \mu\, R\otimes\bar W 
  +(R\otimes\bar W)\mathcal{L}[w](-T)(R\otimes\bar R) + (R\otimes\bar R)\mathcal{L}[w](-T)(R\otimes\bar W)\Big)|r)\bigg],
\label{eq:Ham}
\end{multline}
\end{widetext}
{where $\mathcal{L}[w](-T) = \int_0^\infty dz\, w(z)\, e^{-(-T)z}$ corresponds to the Laplace transform of the interaction potential $w(z)$ of Hamiltonian~({\ref{eq:H}})~{\cite{haegeman_thesis}}.} For our exponentially-decaying interaction potential~(\ref{eq:wexp}), we have $\mathcal{L}[w_{\exp}](-T) = \frac{g\eta}{2}(\eta-T)^{-1}$. In the LL limit, $\mathcal{L}[w](-T)=\frac{g}{2}\mathbbm{1}$, and the term proportional to $\mathcal{L}[w_{\exp}](-T)^2$ in \Eq{eq:Ham} is zero. Taking the derivative of \Eq{eq:min}, now with respect to $Y^{\dagger}$, yields the equation for the optimal $Y^*$,
\begin{equation}
2 \pi \delta(0)\, Y^* = \frac{\delta}{\delta Y^{\dagger}}\langle\Phi|H|\Psi\rangle
\end{equation}
and hence $W^*$ and $V^*$. From these, $Q$ and $R$ are evolved forward in time by a step $d\tau$.

As mentioned above, we fix the gauge of $\ket{\Psi}$ such that $l=\mathbbm{1}$. This is achieved by choosing an arbitrary $R$ and an anti-hermitian $K$, and setting $Q=K-\frac{1}{2}R^{\dagger}R$. It is then possible to construct an update for $Q$ and $R$ that preserves this gauge exactly,
\begin{equation}
\begin{split}
  R(\tau+d\tau)&=R(\tau)-d\tau \,W^*(\tau)\\
  K(\tau+d\tau)&=K(\tau)+\frac{d\tau}{2}\Big(R(\tau)^{\dagger}W^*(\tau) - W^*(\tau)^{\dagger}R(\tau)\Big).
\end{split}
\end{equation}
Indeed one can check that $Q(\tau+d\tau)=K(\tau+d\tau)-\frac{1}{2}R(\tau+d\tau)^{\dagger}R(\tau+d\tau)$. To first order in $d\tau$ equals the update $R(\tau+d\tau)=R(\tau)-d\tau W^*(\tau)$ and $Q(\tau+d\tau)=Q(\tau)-d\tau V^*(\tau)$ (remember $l=\mathbbm{1}$).

For the homogeneous case, it is also possible to manually regauge the matrices $Q,R$ after each update step. Suppose $Q,R$ are ungauged CMPS matrices. Regauging them such that $l=\mathbbm{1}$ is done by first calculating the left eigenvector $(l|$ of $T$ to the eigenvalue $\alpha$ with largest real part. The state $\ket{\Psi}$ is then renormalized by $Q \rightarrow\tilde Q= Q-\frac{\alpha}{2}\mathbbm{1}$. $R$ and $\tilde Q$ are then transformed according to
  $Q_l = l^{1/2}\tilde Q\,  l^{-1/2}$ and 
  $R_l = l^{1/2} R\,  l^{-1/2}$.
It is easy to check that for these matrices, the left eigenvector $l$ with eigenvalue 0 is indeed the identity matrix.

The inverse $(-T)_P^{-1}$ in \Eq{eq:Ham} is to be understood as a pseudo-inverse acting on the vector space
orthogonal to $|r)(l|$. This regularizes the infinite energy content coming from summing up all local energy contributions
to the left of a particular position $x$. The action $(k|(-T)_P^{-1}\equiv(\tilde k|$ on an arbitrary given vector $(k|$ is computed from solving the inhomogeneous linear equation system
$
  (k|\left[\mathbbm{1}-|r)(l|\right]=-(\tilde k|\left[T-|r)(l|\right]
$
for $(\tilde k|$, using a sparse solver~\cite{saad_gmres} for non-hermitian equation system. A similar approach without any pseudo-inverse is used to calculate $(\eta-T)^{-1}$. To get a stable update we have used an implicit Euler scheme to update $Q$ and $R$. The overall operational cost of the procedure explained above is $\mathcal{O}(D^3)$. Depending on the parameters $\eta$ and $g$, time steps $d\tau$ have to be chosen as small as $d\tau=10^{-3}$. 


\begin{thebibliography}{10}
\bibitem{NoteCmps}
The fittings ought to be done in the range $\rho_0 x \gg 1$, where the Luttinger liquid correlations are valid; and $\rho_0 x < \text{const.}\times \log D$, where the CMPS can accurately describe the ground state.
\end{thebibliography}
\end{document}